# Macroscopic Violation of Duality Generated on a Laser Beam

Daniel Mirell[1] and Stuart Mirell[2, *]

[1]*Department of Chemistry, University of California at Irvine, Irvine, California, USA*
[2]*Department of Radiological Sciences, University of California at Los Angeles, Los Angeles, California, USA*
(Dated: July 26, 2013)

It is shown theoretically and experimentally that passage of a laser beam through particular conventional Ronchi gratings prepares the beam in an altered state that violates quantum duality. The violation is characterized by a readily measured net transfer of energy between the prepared beam and an unmodified beam from a similar, independent laser. Notably, the experiment is conducted with the beams at macroscopic power levels where measurability of the duality-violating transfer is vastly improved over that of the discrete photon regime. These results are consistent with other recently reported experiments that have challenged the validity of the duality-related principle of complementarity.

## 1. Introduction

Wave-particle duality is a central principle of the standard probabilistic interpretation of quantum mechanics. We report the results of an experiment violating that principle. The experiment demonstrates a net power transfer between a specially prepared state of a continuous-wave (cw) laser beam $\Phi_G$ and an ordinary beam $\Phi_R$ at the same wavelength but generated by an independent laser where both beams are at macroscopic power levels.

The net transfer is established by first intersecting $\Phi_G$ and $\Phi_R$ over a common coupling path and measuring beam power on a sample region of the $\Phi_G$ beam spot separated from $\Phi_R$ at the terminus of that path. $\Phi_R$ is then blocked from the coupling path and beam power of $\Phi_G$ in the same sample region is re-measured to provide a baseline power for the calculation of the net power transfer.

$\Phi_G$ is prepared in three different states, "depleted", "enriched", and "ordinary", in order to provide for three distinct experimental trials. The total beam power of $\Phi_G$ is substantially identical for these states. The results of these trials show that when $\Phi_G$ is prepared in a depleted state, coupling to $\Phi_R$ results in a net transfer of power to $\Phi_G$. Alternatively, when $\Phi_G$ is prepared in an enriched state, a net transfer of power out of $\Phi_G$ occurs when coupled to $\Phi_R$. Lastly, $\Phi_G$ is prepared in an ordinary state which is shown to yield no net transfer upon coupling with $\Phi_R$.

________
*Electronic address: smirell@ucla.edu

From the perspective of the probabilistic interpretation, independently generated $\Phi_G$ and $\Phi_R$ will still demonstrate interference along an intersection region of those beams, despite Dirac's dictum that a photon can interfere only with itself [1], based upon a theoretical analysis by Mandel [2]. However, the probabilistic interpretation provides no mechanism by which the observed net transfer would occur when $\Phi_G$ is in a presumptive altered state of depletion or enrichment. Specifically, we show here that the source of the duality violation originates with the preparation of $\Phi_G$ in these presumptive altered states.

As an alternative, a theoretical construct derived directly from principles consistent with local realism physically represents these altered states and quantitatively predicts the respective transfers reported here.

## 2. Background

The probabilistic interpretation of quantum mechanics is distinguishable from local realism by the self-interaction of discrete photons incident on mechanisms such as a beam splitter or a double slit. For local realism, the probabilistic property of duality in the discrete photon regime is violated by the wave packet that emerges from one of the two outputs of such mechanisms as an "empty" wave. That empty wave is forbidden by duality which requires an inseparability of a photon's wave-like property, probability, and its particle-like property, the energy quantum. Duality effectively imposes a fixed proportionality between those two properties. In local realism, the treatment of wave-like and particle-like properties as manifestations of two separate and real entities has its origins in de Broglie's initial reality-based pilot wave representation of quanta phenomena [3].

Empty waves, if they exist, should in principle yield testable consequences for a variety of phenomena commonly associated with the wave-like properties of photons. Important advances in the study of these phenomena have been made by a number of investigators including Croca, Garuccio, Lepore, Moreira, Selleri, and others e.g. [4-7]. Popper, in seeking a rational hypothesis for empty waves, has raised compelling philosophical arguments that question the validity of the prevailing probabilistic interpretation [8].

Testability for empty waves in the context of double slit experiments is related to determining which slit is traversed by an energy-bearing photon without destroying accompanying interference arising from both slits, a result consistent with an empty wave traversing the other slit but in conflict with Bohr's duality-related principle of complementarity [9]. Recently, two notable investigations [10] and [11] employing different methodologies have experimentally realized slit traversal determination with verified accompanying interference.

The methodology of one investigation has its origins in an analysis by Aharonov et al. [12] showing that a "weak" measurement of a system can provide some degree of information about the system without significantly altering its subsequent interactions thereby providing an effective investigative tool [13-16]. Wiseman's assessment that such weak measurements could be used to establish average trajectories [17] is experimentally demonstrated in the recent investigation by Kocsis et al. where averaged slit-specific trajectories are measured from a double slit accompanied by interference in the far field of the slits [10].

In a different approach, Rabinowitz proposes a novel double slit experiment using correlated photon pairs in which the passage of discrete photons through a particular slit can be definitively established together with an observation of interference in the far field beyond the slits [18-19]. A recent investigation by Menzel et al. reports the results of an analogous correlated photon pair experiment in which the slit traversed by a discrete photon is determined and far-field interference arising from both slits is verified [11].

The investigations referenced above in this section are inherently associated with the microscopic regime of discrete quantum entities. In this regard it is significant that local realism imposes no constraints on duality violation in the discrete quantum regime as well as in the cw regime where a parameter such as beam power can be at a macroscopic level.

Probability in local realism is a relative entity associated with a real wave structure and retains its original restricted interpretation in Born's rule [20] as the likelihood or "propensity" [8] of a particle-like energy quantum progressing onto a particular probability channel in a complete set of out-going probability channels. A re-scaling of the total probability distributed to the complete set does not alter the likelihood of the particle-like entity entering onto that particular channel. We continue to use the term "probability" here bearing in mind that its usage is potentially misleading since that term suggests equivalence to a mathematical absolute probability. That equivalence is manifested in the probabilistic interpretation as the principle of duality.

In the macroscopic cw regime the treatment of probability as a relative quantity implies that for any number of resident energy quanta on some arbitrary beam segment there is not an imposed, fixed proportionate value of inclusive probability. Consequently, a mechanism predicted by local realism to generate even a modest (duality-violating) disproportion on a cw beam of macroscopic power contributes to conclusive experimental testability. This advantage arises because the net "excess" or "deficit" of particle-like energy quanta residing on the probability's wave-like structure would provide a readily measurable macroscopic power increment.

Clearly, a beam splitter does not provide local realism with the requisite mechanism to achieve duality violation in the macroscopic cw regime. In the transition from the discrete to the macroscopic regimes as beam power is increased, statistical distribution of the numerous incident energy quanta onto the beam splitter's two output channels restores a proportionality of probability and energy quanta on both of those channels that is not distinguishable from that of duality. However, there are a number of mechanisms consistent with local realism, e.g. [21], that generate duality violation at macroscopic power levels. We report here on one of those mechanisms, its theoretical basis, and on the experimental demonstration of duality violation using that mechanism.

## 3. Mechanism for Duality Violation

The basic mechanism for duality violation presented here can most readily be appreciated by first considering a simple gedanken experiment before proceeding to the particular realization of that experiment reported here. In the gedanken experiment we use a Ronchi transmission grating (ruling) and a laser beam at normal incidence. A

Ronchi grating consists of a linear array of equal-width opaque and transmissive bands resulting in a grating period $p=2w$ where $w$ is the "slit" width represented by the transmissive bands. In the gedanken experiment, we hold the laser beam wavelength constant at some $\lambda$ and consider the consequences of progressively decreasing $w$ from large values of $w \gg \lambda$ down to small values of $w \sim \lambda$ while maintaining $p=2w$.

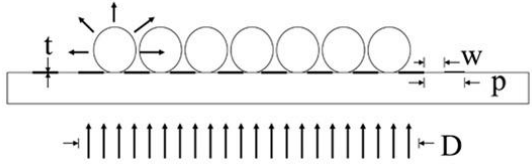

**Figure 1.** Emergent output diffraction wavelets from irradiated individual slits fully determine collective output probability (and energy) prior to the formation of resultant orders by interference as those wavelets intersect.

The irradiated slits produce the **Figure 1** "output" set of wavelets that do not intersect in the grating's near field. The total output probability emerging on these wavelets is a constant $P_o$ independent of $w$ since the transmissive factor for a Ronchi is $0.5=w/p$. In arbitrary units, we can set $P_o=1$. As the individual wavelets expand in the grating's far field, they generate smoothly-varying single-slit envelopes that intersect and interfere thereby producing the "resultant" set of highly directional diffraction orders $j$ shown in **Figure 2** for the Fraunhofer approximation. The envelope intensity is proportional to $\sin^2\alpha/\alpha^2 = \mathrm{sinc}^2\alpha$ where $\alpha=(\pi w/\lambda)\sin\theta$ and $\theta$ is the physical azimuthal angle. The depicted envelope in **Figure 2** is equivalent to an individual slit envelope scaled up by the total number of irradiated slits to correctly show the essential equivalence of the output integral and the integral over the resultant peaks. For Ronchi gratings, the side lobes are bisected by the symmetrically disposed odd $\pm j$ orders for $|j|$ and the integral over a particular side lobe is equal to the integral of the odd $j^{th}$ resultant peak bisecting that side lobe. **Figure 2** is parameterized in terms of the continuous variable $j=2\alpha/\pi$ as well as $\alpha$ since $j$ conveniently assumes integer values at the side lobe bisector points and null points.

Before proceeding further, we introduce a convention useful in the present theoretical analysis. Beam probability $P$ is evaluated by volumetric integration of intensity over a beam segment of some arbitrary selected length. Correspondingly, this defines an inclusive energy $E$ on that beam segment. We use the convention of considering beam segments purely as a convenience in order to proceed with the theoretical analysis using the parameters of probability and energy rather than the associated flux densities of those parameters.

Nevertheless, the results of the analysis are fully applicable to the experimentally relevant parameter of the integrated energy flux density, i.e. power, which is proportionate to that inclusive energy $E$.

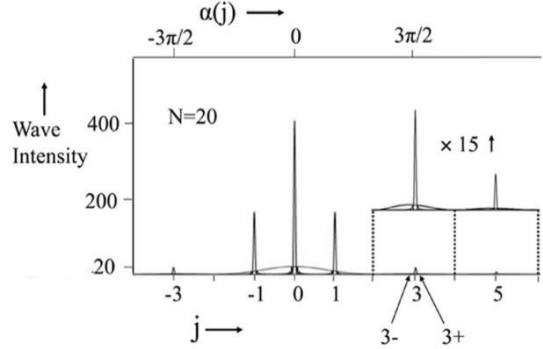

**Figure 2.** Single slit output diffraction intensity envelope and associated resultant diffraction intensity peaks for a Ronchi grating of those slits. The single slit envelope is scaled up by number of irradiated slits N=20 to correctly depict total envelope output intensity relative to that of the resultant peaks. The depicted total envelope output and the resultant peaks are both scaled up by ×15 in the detail graph of the $j=2$ to 6 range.

From a straight-forward classical calculation [22], the resultant probability of the diffraction orders

$$P_r(j_t)_F = \frac{\dfrac{\pi}{2}\sum_{j=-n}^{n} \mathrm{sinc}^2\alpha_j}{\displaystyle\int_{-\alpha_t=-\pi j_t/2}^{\alpha_t=\pi j_t/2} \mathrm{sinc}^2\alpha\, d\alpha} \quad (1)$$

is given by the integral over the resultant order intensities, expressed here as a Riemann sum, divided by the integral of the output envelope intensity. The output envelope truncates at $\alpha_t=\pm\pi w/\lambda$ where the azimuthal $\theta=90°$. Truncation on the $j$ continuum is given by $j_t=2\alpha_t/\pi$. $P_r(j_t)_F$ is shown as a function of the convenient parameter $j_t$, but that functional dependence is fundamentally the linearly related parameter $\alpha_t$. $P_r(j_t)_F$ is subscripted by F to indicate that the quantity is valid in the Fraunhofer approximation. Because of symmetry, dependence on $j_t$ is equivalent to that for $-j_t$. From these expressions, we have a useful relationship for the Ronchi slit width in terms of $j_t$,

$$w = j_t\lambda/2. \quad (2)$$

The Riemann sum truncates at the $\pm n^{th}$ final resultant orders inclusive within $\pm\alpha_t$. Those $\pm n^{th}$ orders are respectively in the neighborhood of $\theta=\pm90°$ and are said to be near the grating "threshold".

The basis for the derivation of Equation (1) [22] can be succinctly summarized here. An integral over resultant order intensities (or its Riemann sum equivalent) would normally by itself constitute the requisite resultant probability unlike the form of Equation (1). However, in the present case the integral over resultant orders as well as the output envelope integral are both computed over $\alpha$ and not the physical azimuthal angle $\theta$ causing both integrals to exhibit an artifactual decrease as the variables $\alpha_t$, $j_t$, and $w$ in our gedanken experiment mutually decrease. In particular, the total output probability $P_o$ was shown above to physically remain constant independent of $w$. Accordingly, the divergence from constancy of the output integral evaluated over $\alpha$ yields the requisite normalization factor for the corresponding resultant order integral over $\alpha$ to provide a probability $P_r$ for which the $\alpha$-related artifactual decrease is selectively eliminated.

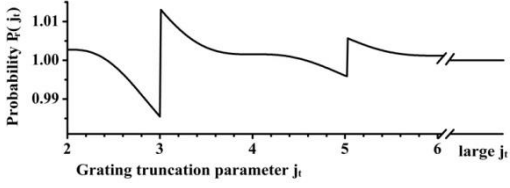

**Figure 3.** Total resultant probability $P_r(j_t)$ inclusive of all diffraction order probabilities within the truncation limits $\pm j_t$. For large $j_t$, the total resultant probability $P_r(j_t)_F = 1$ is valid.

Turning now to the beginning of our gedanken experiment in the limit of large $w$, which corresponds to "coarse" gratings, the significant central region of the output envelope and all of the correspondingly significant resultant diffraction orders are tightly clustered in the forward direction about 0° and, equivalently, about $\alpha=0$ consistent with the Fraunhofer approximation. The more distal envelope and higher orders extending out to $\pm 90°$ are vanishingly small where both "truncate" at some large $|j|$. In this limit the Equation (1) $P_r(j_t)_F$ is very nearly a constant value of unity. This constancy for large $w$ merely expresses the high degree to which probability is conserved for a coarse grating in the transition that occurs as the expanding wavelets interfere and generate in their place the highly directional resultant orders.

However, as $w \to \lambda$ for "fine" gratings, only the envelope's central maximum and significant proximal side lobes are included over the $-90°$ to $+90°$ full azimuthal span as truncation occurs at some small $|j|$. Correspondingly, that same truncation also leaves only a very few orders, including those with the most significant intensities, widely dispersed over $-90°$ to $+90°$. This departure from the Fraunhofer approximation necessitates that we amend the Equation (1) $P_r(j_t)_F$ to properly represent the relative intensities of the resultant orders as a function of inclination from 0° by incorporating the classically well-known "obliquity" correction appropriate for the presently considered Ronchi grating diffraction system. The relative actual intensities of orders nearer threshold are reduced below that predicted by the $sinc^2 \alpha_j$ function. A physically accurate obliquity correction in the present experiment is readily available empirically from a direct measurement of the beam powers of the resultant $j^{th}$ orders emerging from an appropriate fine grating and comparing those powers to the $sinc^2 \alpha_j$ function. Our primary interest in this investigation is confined to gratings for which $j_t$ ranges from $\pm 4$ down to $\pm 2$ respectively over the first side lobes. For a Ronchi grating the probabilities (integrated intensities) of those output side lobes are expressed in turn by the probabilities associated with the bifurcating resultant $\pm 3^{rd}$ orders. Accordingly, we measure the powers of the resultant orders with a Ronchi grating that places the $\pm 3^{rd}$ orders near threshold. From these measurements, we obtain an obliquity correction of $f=0.56$ for those $\pm 3^{rd}$ orders near threshold and a correction negligibly differing from unity for the lower orders within the diffraction envelope's central maximum. Similarly, that obliquity correction $f$ is also applicable to the output integral over the first side lobe.

The obliquity-amended expression for the Equation (1) resultant probability

$$P_r(j_t) = \frac{0.5\pi(0.5\,sinc^2\alpha_{j=0} + sinc^2\alpha_{j=1})}{\int_0^\pi sinc^2\alpha\,d\alpha + f\int_\pi^{\pi j_t/2} sinc^2\alpha\,d\alpha}$$
$$+ \frac{0.5\pi f\, S_3\, sinc^2\alpha_{j=3}}{\int_0^\pi sinc^2\alpha\,d\alpha + f\int_\pi^{\pi j_t/2} sinc^2\alpha\,d\alpha} \quad (3)$$
$$= \frac{2.539 + 0.071\, S_3}{2.532 + \int_\pi^{\pi j_t/2} sinc^2\alpha\,d\alpha}$$

is valid when the truncation $j_t$ resides in the first side lobe ($2 \leq j_t \leq 4$). Because of symmetry about 0 on the $j_t$ continuum, negative $j$ terms have been excluded in the Riemann sum. Since the negative $\alpha_t$ limits in the envelope integration have also been excluded, Equation (3) still represents the total resultant probability for all positive and negative orders within $\pm j_t$. $S_3$ is a step function, zero when the $j=3$ order is excluded and unity when the $j=3$ order is included in the integrated output envelope. $P_r(j_t)$ exhibits a discontinuous perturbation about unity at $j_t=3$. This perturbation is also present for the other odd $j_t>3$, but these perturbations rapidly diminish with increasing $j_t$ as depicted in **Figure 3**. In this figure, the obliquity correction is applied to the second side lobes as well as the first side lobes in order to depict the diminution of the perturbation at $j_t$=5 relative to that at $j_t$=3. The perturbations are not true mathematical discontinuities but approach that status as the number of irradiated slits becomes large.

The origin of these perturbations is understood from Equation (3). Interference of the **Figure 1** wavelets converts the output probability into resultant probabilities represented by the highly directional diffraction orders. As $w \rightarrow \lambda$, truncation leaves only a very few orders propagating over −90° to +90°. Consider for example the behavior of the Equation (3) probability beginning at $j_t$=4, i.e. $P_r(4)$. Truncation at $j_t$=4 includes the envelope's central maximum and the two entire adjacent side lobes. The only non-zero propagating orders are the $0^{th}$, $\pm 1^{st}$, and $\pm 3^{rd}$. $P_r(4)$=1.0015 ≈1 shows that probability is substantially conserved in the output→resultant transition at the $j_t$=4 truncation value. As we progress from $j_t$=4→$3^+$ (where the notation $3^+$ designates marginal inclusion of the $\pm 3^{rd}$ order peaks), the intensities of the propagating orders remain constant relative to each other as seen from the Riemann sum in the numerator of Equation (3). However, concurrently the normalization factor provided by the output envelope integral over $\alpha$ progressively decreases as truncation reduces the first side lobes. In this process as $j_t$=4→$3^+$, output wavelet interference generates a progressively increasing resultant probability $P_r(j_t)$ to about 1.3% above unity.

Most significantly for our gedanken experiment, in the incremental truncation point reduction $j_t$=$3^+$ →$3^-$ the $\pm 3^{rd}$ orders are abruptly excluded from the Riemann sum. The resultant probability $P_r(j_t)$ exhibits a nearly discontinuous fall of about 2.8% ending at 1.5% below unity as the normalizing integral of the output envelope is virtually unaltered in that incremental $j_t$=$3^+$ →$3^-$ reduction. (**Figure 2** necessarily shows resultant orders formed from only 20 irradiated slits to clearly depict the envelope on the same vertical scale. For an experimentally realistic N~500 irradiated slits, the base widths of the $\pm 3^{rd}$ order peaks are a mere ~0.2% of the respective side lobe widths emphasizing the incremental nature of the $j_t$=$3^+$ →$3^-$ transition that removes a significant pair of resultant probability channels with a vanishingly small change in the normalizing integral.)

Over the subsequent transition $j_t$=$3^-$ →2, interference of the individual output envelopes continues to decrease the normalization integral in Equation (3). $P_r(j_t)$ increases asymptotically to $P_r(2)$=1.0028≈1 and at $j_t$=2 probability is again substantially conserved in the input→output transition.

These excursions of resultant probability $P_r(j_t)$ are still, by themselves, unremarkable. Classically, there is no inherent violation associated with the interference of an initial wave set producing a final wave set where the respective integrated wave intensities of each (identified in local realism as relative probabilities) may differ as a result of net destructive or constructive interference. Correspondingly for local realism, relative probability is not necessarily a conserved quantity.

The origin of duality violation relates to the associated energy output of the grating. Each individual slit irradiated by an ordinary incident beam (**Figure 1**) produces an output probability sampling that is accompanied by a proportionate energy sampling. Summation over all irradiated slits yields collective output quanta with energy $E_o$ proportionate to the collective output probability $P_o$ which was earlier assigned a unit value. Because of the proportionate samplings of these two output quantities, in dimensionless units $E_o$ may also be set to unity giving

$$P_o = 1 = E_o. \quad (4)$$

The equality of the wave-like probability and the particle-like energy is effectively a statement that the output of the grating in the near field is still in agreement with duality. This concurrence with duality can be formalized by defining an "occupation" value $\Omega$ as the ratio of resident energy quanta on a wave of some probability. For the grating output, the occupation value is

$$\Omega_o = \frac{E_o}{P_o} = 1. \quad (5)$$

The critical observation to be made here concerns the distribution of the output quanta represented by $E_o$ that initially reside on the emergent near-field wavelets. As the wavelets expand and intersect, the resident energy quanta distribute without loss onto the resultant $P_r(j_t)$ far-field probability channels (the diffraction orders) in

proportion to the respective relative probability channels. The resultant energy

$$E_r = E_o = 1 \qquad (6)$$

and is fully conserved for all values of $w$ and, equivalently, of $j_t$. The resultant occupation value

$$\Omega_r(j_t) = \frac{E_r}{P_r(j_t)} = P_r(j_t)^{-1} \qquad (7)$$

is simply the inverse of the resultant probability as shown in **Figure 4** and, by symmetry, is equally applicable to any $\pm j_t$. $\Omega_r(j_t)$ is re-identified as the theoretically predicted $\Omega_{G\,th}$ on that graph. Because of non-conservation of probability, there are regions with $\Omega_r(j_t)>1$ where the resultants are defined as "enriched" and regions with $\Omega_r(j_t)<1$ where the resultants are defined as "depleted" in reference to the respective disproportionalities of resident energy quanta relative to probability.

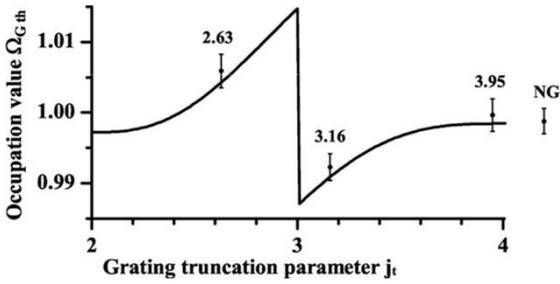

Figure 4. Theoretical occupation value curve $\Omega_{G\,th}$, equivalent to the Equation (7) $\Omega_r(j_t)$. The three experimentally measured $\Omega_{G\,ex}$'s are plotted for comparison to the theoretically predicted $\Omega_{G\,th}$. The no-grating NG control experiment value $\Omega_{NG\,ex}$ is plotted to the right relative to the vertical axis.

Resultant beams with $\Omega=1$ are defined as "ordinary", a designation that also applies to the incident beam and the grating output. In the realization of our gedanken experiment, we use particular Ronchi gratings with $w$ values that are predicted to respectively generate enriched and depleted resultant beams. Any individual $j^{th}$ order of the resultant orders has the same occupation value as that of the entire set, i.e. $\Omega_{rj}=\Omega_r$, because occupation values are intrinsic variables formed from quotients of the extrinsic variables of energy and probability. A particular $j^{th}$ order beam with a probability $P_{rj}$ acquires a $P_{rj}/P_r$ share of the total energy $E_o=E_r$ as it leaves the near-field of the grating.

Before proceeding with the performed experiment, we note that the discontinuity encountered in a transition such as that from the depleted region to the enriched region when $j_t=3^+ \rightarrow 3^-$ is characterized by the energy quanta that had been on the $\pm 3^{rd}$ orders being redistributed onto the remaining propagating orders. This abrupt redistribution of energy onto remaining orders as an order passes threshold is superficially analogous to a "Rayleigh grating anomaly". Historically, the incident beam used to study grating anomalies is multi-wavelength where the beam energy (or power) is at most a slowly varying function of wavelength. Rayleigh anomalies relate to those photons at some particular wavelength $\lambda_t$ that have an order "at threshold" i.e. at the grating plane. In Rayleigh's analysis [23], the photons on that threshold order are coherently scattered off of deep grating grooves (relative to $\lambda_t$) and are redistributed onto the remaining propagating orders of the $\lambda_t$ photons. As a result, those remaining orders of the $\lambda_t$ photons exhibit an abrupt increase in energy relative to the energy on the corresponding orders of marginally shorter wavelength $\lambda_{nt}$ photons that have an unscattered order near threshold (denoted by "nt") but not at threshold. Significantly, the scattered $\lambda_t$ photons carry not only the energy quanta but also the associated wave packet onto the remaining propagating orders. Consequently, the Rayleigh anomaly is consistent with probabilistic duality (as well as with local realism). In this context, we note that the Ronchi gratings used in this experiment have thin opaque bands relative to wavelength (see **Figure 1** where "t" refers to thickness), and would not be expected to provide the photon scattering mechanism associated with gratings actually demonstrating the Rayleigh anomaly. Clearly, a methodology to directly measure the duality state of the Ronchi grating propagating orders must be utilized in order to establish whether those orders are ordinary, as would be expected for Rayleigh anomalies, or are in fact "duality-modulated" ($\Omega$ deviations from unity) as predicted by local realism.

The experiment presented in the next section to provide that direct test for duality modulation utilizes a transient coupling between a resultant presumptively duality-violating beam from the grating (a propagating order) and an independent ordinary beam [22]. The coupling setup is analogous to the intersection of two independent beams used in numerous investigations to experimentally assess duality violation by determining the presence or absence of interference between the intersected beams as a test of the probabilistic interpretation. An excellent review of these investigations is given by Paul [24]. An observation of interference would seemingly violate Dirac's dictum that a photon in the probabilistic interpretation can interfere only with itself [1].

The outcomes of these numerous investigations are conclusive demonstrations that interference does occur in apparent contradiction to the probabilistic interpretation. However, in a theoretical analysis of this phenomenon, Mandel makes the critical argument that for any given photon measured in the interference we do not know on which beam that photon had initially resided [2]. Because of that lack of knowledge, each photon is treated in Mandel's analysis as interfering with itself. Consequently, interference in the intersection of independent beams is widely regarded as consistent with the probabilistic interpretation and as not providing a test of that interpretation.

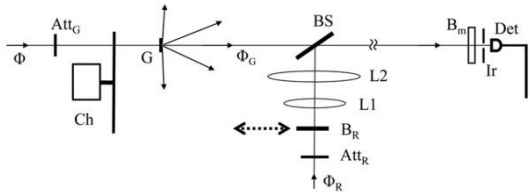

**Figure 5.** Experimental apparatus configuration showing a potentially duality-modulated beam $\Phi_G$ after passage through a particular grating $G$. $\Phi_G$ is equilibrated with an ordinary beam $\Phi_R$ along a coupling path extending from beam splitter BS. Coupling occurs with beam blocker $B_R$ shifted to transmit $\Phi_R$.

In a variant of those independent beam investigations, we prepare one of the beams by transmission through a particular grating where a prepared beam $\Phi_G$ may specifically violate probabilistic duality, i.e. the beam is in a depleted or enriched state from the perspective of local realism. Spatially transient coupling of that prepared beam with an independent ordinary "restoration" beam $\Phi_R$ by mutual interference over a coupling path should then yield a net equilibrating transfer of energy quanta for local realism but not for the probabilistic interpretation. That net energy transfer relative to $\Phi_G$ is experimentally readily measurable in the cw regime at macroscopic powers by detecting the beam power on a sampling of $\Phi_G$ that is substantially separate from $\Phi_R$ at the end of the coupling path with and without $\Phi_R$ present on the coupling path.

## 4. Experimental Configuration and Methods

We begin with a description of the experimental configuration shown in **Figure 5**. A HeNe laser generates a horizontally linearly polarized beam $\Phi$ of several milliwatts at 633 nm. Beam $\Phi$ traverses a variable attenuator $Att_G$ and an optical beam chopper wheel Ch with a 0.5 duty cycle generating square wave pulses at 40 Hz. $\Phi$ is at normal incidence on a grating $G$. The grating is one of three Ronchi gratings (rulings) with respective slit widths $w$=833 nm, 1000 nm, and 1250 nm formed from opaque bands of 150 nm-thick reflective chromium deposited on a glass substrate. The respective grating frequencies in lines/mm are 600, 500, and 400. The slit widths uniquely characterize each grating through Equation (2) which gives the $w$-dependent $j$-continuum truncation points in terms of the slit widths

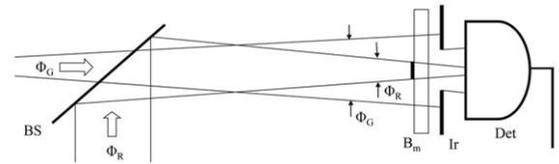

**Figure 6.** Detail of Figure 5 coupling path, not to scale, showing the substantial separation of $\Phi_R$ from $\Phi_G$ at the detector Det by convergence of the former onto a disk mask beam stop $B_m$.

where $\lambda$=633 nm in this investigation. These points are

$$j_t(w) = 2.63, 3.16, \text{ and } 3.95 \qquad (8)$$

for the three respective slit widths. Accordingly, the individual gratings are also uniquely characterized by the Equation (8) $j$-continuum truncation points. In our notation a Ronchi grating of some arbitrary slit width $w$ is denoted as $G(j_t)$ or simply $G$. Conversely, a grating identified with a numerical $j_t(w)$-equivalent truncation point identifies a particular grating with an implicitly expressed slit width as in $G(2.63)$, $G(3.16)$, and $G(3.95)$ for the three selected gratings. This use of $j_t$ continues our convention in which variables are most instructively identified by the critical $j$-continuum truncation value.

The motivations for selecting these three gratings are their duality properties based on Equation (7). $G(2.63)$,

$G(3.16)$, and $G(3.95)$ are respectively predicted to produce enriched, depleted, and ordinary resultants. Data are also acquired with no grating present. These trials, which are designated by NG, should produce ordinary resultants and serve as control experiments.

The particular grating under study is mounted with the grating bands on the exit face and vertically oriented thereby providing (transverse magnetic) TM (S) polarization with respect to $G$ in the usual classical configuration for observing grating anomalies. The $0^{th}$ order diffraction beam identified as $\Phi_G$ is incident on a 50:50 metallic plate beam splitter BS with a 3 mm thick glass substrate and the metallic deposition on the exit face. An independent HeNe laser generates a horizontally linearly polarized beam $\Phi_R$ initially several milliwatts in power. $\Phi_R$ traverses a variable attenuator $Att_R$, a retractable beam blocker $B_R$ and enters a beam expander, L1 ($f=+100$ mm) and L2 ($f=+200$ mm), before forming a beam spot concentric with that of $\Phi_G$ on the beam splitter BS as shown in the **Figure 6** detail view. This concentricity is a critical alignment for the apparatus.

Measurement of duality violation with this **Figure 5** configuration imposes some essential general criteria on the set of beam and apparatus parameters in the region designated as the "coupling path" that extends from BS to the final optical components. We include those general criteria below, augmented by specific examples of parameter values given in parentheses that are taken from an experimentally utilized parameter set. In that set, specified beam widths are Gaussian diameters.

The equivalency of the $\Phi_G$ and $\Phi_R$ polarization angles should be verified on the coupling path itself and corrected by rotation of $\Phi_R$ if necessary. At BS the diameter of $\Phi_R$ (1.8 mm) is expanded slightly beyond that of $\Phi_G$ (1.7 mm) as a result of the L1 and L2 relative spacing. The beam components exiting BS utilized here are the transmitted component of $\Phi_G$ and the reflected component of $\Phi_R$. The orientation of BS is adjusted to concentrically align the $\Phi_R$ beam spot to the $\Phi_G$ beam spot at the terminus of the coupling path. This critical, second beam spot alignment coaxially aligns the $\Phi_R$ and $\Phi_G$ beams over the coupling path length (~2000 mm).

An assembly of a beam blocker $B_m$, an iris diaphragm Ir, and a photodiode detector Det is located at the coupling path terminus. $B_m$ consists of an opaque disk mask beam stop (diameter 1.7 mm) mounted on a glass substrate as detailed in **Figure 6**. At $B_m$, natural divergence from the source laser has further increased the diameter of $\Phi_G$ (3.8 mm) to a value significantly larger than that of the $B_m$ disk mask. Conversely, the relative spacing of L1 and L2 is critically adjusted to converge the $\Phi_R$ diameter (1.0 mm) to a value significantly less than that of the $B_m$ beam mask. The iris (3.3 mm dia.) is set to exclude the peripheral portion of $\Phi_G$ from the detector Det. Beam directors on the coupling path (not shown in **Figures 5** and **6**) are used to provide concentric alignment of $\Phi_G$ and $\Phi_R$ with $B_m$, Ir, and Det.

Data are acquired with one of the three Ronchi gratings in the **Figure 5** position of $G$. We continue the use of $\Phi_G$ and $\Phi_R$ as general identifiers of the respective beams, but we are reminded that these wave functions in local realism are exclusive of the energy quanta residing on those beams.

Consistent with our prior notation, the complete expression for occupancy of that $\Phi_G$ beam would formally be identified as $\Omega_{r0}(j_t)$ since $j_t$ designates the unique truncation point on the $j$-continuum for any Ronchi grating $G=G(j_t)$, $r$ denotes that $\Phi_G$ is a resultant diffraction beam and 0 specifies its order. These identifiers are already understood in the present context and the compact expression $\Omega_G$ is used here in place of $\Omega_{r0}(j_t)$. Similarly, an $\Omega_R=1$ designates the occupation value of the initially ordinary restoration beam $\Phi_R$.

The basic premise of beam coupling is that a duality modulated beam equilibrates with an ordinary beam by a net transfer of energy quanta that leaves the wave structures of both beams unchanged and ideally converges the occupation values toward a common value. For the objective of achieving complete equilibration,

$$\Omega_{Gc} = \Omega_{Rc} \qquad (9)$$

as the two beams $\Phi_G$ and $\Phi_R$ approach the end of the coupling path. We use the added subscript "$c$" on variables such as $\Omega$ to denote values at the end of the coupling path where equilibration of $\Phi_G$ and $\Phi_R$ has potentially altered those values (in contrast to the respective values of those variables without $\Phi_G$ and $\Phi_R$ simultaneously present on the coupling path). $\Phi_R$ should ideally serve as an infinite source for a depleted $\Phi_G$ or an infinite sink for an enriched $\Phi_G$ in the equilibration process. This criterion is satisfactorily approximated when the inequality

$$P_R \gg P_G \qquad (10)$$

is satisfied by a ratio of ~100:1 leaving the final equilibrated $\Phi_G$ and $\Phi_R$ both as ordinary and extending the Equation (9) $\Omega$ equality to a unit-valued ordinary value,

$$\Omega_{Gc} = \Omega_{Rc} = 1. \qquad (11)$$

If $\Phi_G$ is depleted or enriched as it emerges from a grating $G$, a net transfer of energy $\Delta E$ will occur between $\Phi_G$ and $\Phi_R$ that changes the initial grating-emergent energy $E_G$ of $\Phi_G$ to

$$E_{Gc} = E_G \pm \Delta E \qquad (12)$$

as the coupling path terminus is approached. A positive-signed $\Delta E$ corresponds to an energy gained by $\Phi_G$ in a transfer from $\Phi_R$ where $\Phi_G$ had initially been depleted. Similarly, if $\Phi_G$ had initially been enriched, $\Delta E$ is negatively signed. Alternatively, if $\Phi_G$ emerging from $G$ is initially ordinary, no net transfer occurs and $\Delta E$ is zero.

The coupling equilibration of $\Phi_G$ to an ordinary state (if it is not already in an ordinary state) provides the important result

$$P_G = E_{Gc} \qquad (13)$$

at the coupling path terminus for our choice of arbitrary units. A sampling of this $E_{Gc}$ is acquired by the detector.

Ideally, the wave-like probability $P_R$ is entirely confined to the mask of $B_m$ as $P_{Rm}=P_R$ and the residual probability of $P_R$ in the detector's annular sampling region $P_{Ra}=0$, but this extreme criterion is impractical for the **Figure 5** apparatus. However, this criterion can be satisfactorily approximated by

$$P_{Rm} \gg P_{Ra}. \qquad (14)$$

The criteria given by the inequalities of Equations (10) and (14) are expressed in terms of probabilities which are not directly amenable to measurement. Nevertheless, for the modest deviations of $\Omega$ from unity on the order of one percent realized in the present experiment, the two inequalities are equivalently representable in terms of the corresponding energies $E_R$, $E_G$, $E_{Rm}$, and $E_{Ra}$. These energies, in turn, are proportional to the experimentally measurable corresponding beam powers $PWR_R$ (~600 μW), $PWR_G$ (~6 μW), $PWR_{Rm}$ (~600 μW), and $PWR_{Ra}$ (~2.5 μW) which can be substituted for the probabilities in the two inequalities for the purposes of setting up the apparatus. Correspondingly, the $\Phi_G$ pulse height is measured from the power $PWR_{Ga}$ (~2.5 μW) incident on the detector. (With the chopper wheel in rotation, the detector amplifier registers an average power of $PWR_{Ga}/2$ for the pulsed $\Phi_G$.) The beam power objectives on the coupling path are achieved by adjustment of the variable attenuators $Att_G$ and $Att_R$.

For data acquisition in the experiment with a particular $G=G(j_i)$ in place, the output voltage of the detector amplifier provides a proportionate instantaneous measure of the energy incident on the detector in the annular sampling region. With the chopper wheel in rotation, the detector amplifier signal received by a digital oscilloscope produces a square wave of a height proportional to the pulsed $\Phi_G$ energy.

A trapezoidal-like deviation from a true square wave is caused by partial eclipsing of the $\Phi_G$ beam by the vanes of the chopper wheel. That deviation, which represents incomplete detection sampling of $\Phi_G$, is reduced to an insignificant level relative to the complete sampling when the ratio of the Gaussian diameter of $\Phi_G$ to the arc span between adjacent vanes is very small (~1:100 in the present apparatus). This reduction is most readily achieved when using a chopper wheel with a minimal number of vanes.

With the oscilloscope set to dc coupling of the signal input, the $\Phi_G$ square wave rides on the baseline bias level produced by any steady state flux of photons in the annular sampling region. When $\Phi_R$ is blocked from the coupling path, that flux consists only of background photons. When $\Phi_R$ is unblocked, that steady state flux additionally includes $\Phi_R$ photons residually in the annular sampling region and accordingly significantly elevates the baseline bias level. Conversely, with the oscilloscope set to ac coupling of the signal input, the $\Phi_G$ square wave appears as alternating positive and negative half-height square pulses symmetrically distributed about the horizontal zero voltage midline independent of a blocked or unblocked $\Phi_R$. For either mode of signal coupling, incremental changes in the full pulse height are readily measured as $\Phi_R$ is blocked and unblocked.

With $\Phi_R$ unblocked, i.e. coupled to $\Phi_G$, the pulse height of the square wave, measured as a differential voltage between the upper and lower levels,

$$\Delta V_{Gc} = \kappa E_{Gc} = \kappa P_G \qquad (15)$$

gives the $\Phi_G$ post-coupled energy $E_{Gc}$ and, very importantly, the probability $P_G$ as well to within a multiplicative constant $\kappa$.

After $\Delta V_{Gc}$ is acquired, $\Phi_R$ is blocked from the coupling path by $B_R$. The detector then samples the same annular region of $\Phi_G$ but now the pulse height measurement

$$\Delta V_G = \kappa E_G \qquad (16)$$

provides the beam's grating-emergent energy $E_G$, unmodified by coupling, to within the same multiplicative constant $\kappa$.

The measurement precision of $\Delta V_{Gc}$ and of $\Delta V_G$ is optimized by using real-time waveform averaging. Moreover, the measurement precision of $\Delta V_{Gc}$ relative to $\Delta V_G$ is significantly improved by using ac coupling rather than dc coupling of the input signal. With ac coupling, as $\Phi_R$ is alternately blocked and unblocked, the respective upper and lower levels of the square wave each shift by only ~1 % or less relative to the full pulse height of the square wave. The confinement of the respective upper and lower levels to these narrow ranges of the vertical measurement scale effectively eliminates any artifactual effects of range-related non-linearity.

Conversely, for dc coupling as $\Phi_R$ is unblocked the square wave shifts upward on the vertical measurement scale by the bias voltage produced when a power $PWR_{Ra}$ is deposited on the detector. Consequently, dc coupling necessitates an extremely linear response by the oscilloscope in order to distinguish an actual transfer-related incremental change of $\Delta V_{Gc}$ relative to $\Delta V_G$ from an artifactual change arising from a non-linear response over the full range being used. (For example, with $PWR_{Ra} \sim PWR_{Ga}$, unblocking $\Phi_R$ shifts the waveform on the vertical measurement scale by a bias that is ~100% of the square wave pulse height and the full utilized range in assessing $\Delta V_{Gc}$ relative to $\Delta V_G$ is about two orders of magnitude greater for dc coupling than for ac coupling.)

The vital significance of the $\Delta V_{Gc}$ and $\Delta V_G$ pulse height measurements is that their ratio

$$\frac{\Delta V_G}{\Delta V_{Gc}} = \frac{E_G}{P_G} = \Omega_{G\,ex\,i} \qquad (17)$$

which is the experimentally determined occupation value from a sequential pair of measurements $\Delta V_{Gc}$ and $\Delta V_G$, each of which is derived from an averaging over 128 pulse cycles. The subscript "$ex$" has been added to clearly identify this quantity as experimentally determined. The subscript "$i$" denotes $\Omega_{G\,ex\,i}$ as a single trial value for the particular installed grating $G$.

A typical incremental change in $\Delta V_{Gc}$ relative to $\Delta V_G$ for two sets of these averaged pulses may be on the order of $\pm 1\%$ (for $G(3.16)$ or $G(2.63)$, respectively). For beams of macroscopic power, these incremental changes are representative of enormous numbers of photons (~$10^{11}$) added to or subtracted from $\Phi_G$ in the annular sampling region as $\Phi_R$ is coupled to $\Phi_G$. These changes are directly observable on the oscilloscope waveform as $\Phi_R$ is blocked and unblocked. The oscilloscope also generates digital values for $\Delta V_{Gc}$ and for $\Delta V_G$ from which each single trial value $\Omega_{G\,ex\,i}$ is calculated for the particular installed grating $G$. Ten single trial values $\Omega_{G\,ex\,i}$ with that particular $G$ are acquired and averaged to give a final reported value $\Omega_{G\,ex}$ for that set of individual trials. The process is repeated for the other two gratings to give the three trial-averaged values $\Omega_{G(2.63)\,ex}$, $\Omega_{G(3.16)\,ex}$, and $\Omega_{G(3.94)\,ex}$. An additional set of ten trials is acquired with no grating (NG) present which provides a control experiment value $\Omega_{NG\,ex}$. The resultant beam $\Phi_{NG}$ is ordinary and should exhibit no net transfer upon coupling. In these NG trials, a fixed-value attenuation filter is substituted for the gratings at position $G$ in the apparatus. For any given setting of the variable attenuator $Att_G$, that fixed-value filter transmits approximately the same power to the coupling path as do the gratings. The four sets of trials comprise a complete set.

All of the single trials in the complete set, including the NG trials, are acquired with the same beam power $PWR_G$ on the coupling path. This is facilitated by minor adjustment of attenuator $Att_G$ prior to acquiring each of the four sets to maintain some selected beam power $PWR_{Ga}$ at the detector. The objective of this procedure is to exclude any artifactual power-related influences on the measurements contributing to the determinations of the three $\Omega_{G\,ex}$ and the $\Omega_{NG\,ex}$.

As a practical matter, the apparatus is most readily initially aligned with either grating $G(2.63)$ or $G(3.16)$ installed. Optimum coaxial alignment of $\Phi_G$ and $\Phi_R$ on the coupling path should be observable in real time from a ~1% change in pulse height as $\Phi_R$ is alternately blocked and unblocked. It is important to note that deficiencies in fully achieving the various beam parameter criteria and accurate concentric alignment of $\Phi_G$ and $\Phi_R$ on the coupling path result in an incomplete equilibration transfer and a resultant experimental underestimate of the actual magnitude of the duality modulation [22].

## 5. Experimental Results

The experimentally determined $\Omega_{G\ ex}$ values specific to the three gratings, $\Omega_{G(2.63)\ ex}$, $\Omega_{G(3.16)\ ex}$, and $\Omega_{G(3.95)\ ex}$, are plotted on the **Figure 4** theoretically predicted $\Omega_{G\ th}$ (equivalent to $\Omega_r(j_t)$ from Equation (7) and inclusive of an empirical obliquity correction). The no-grating control experiment value $\Omega_{NG\ ex}$ is also depicted for comparison.

With the 600 line/mm grating $G(2.63)$ installed in the apparatus, the experimentally determined occupation value $\Omega_{G(2.63)\ ex}=1.0059\pm0.0024$ which, alternatively expressed as a "duality modulation" (deviation of $\Omega$ from unity), is $+0.59\pm0.24\%$. This result and the results below are given with $\pm$SE standard error for n=10 trials. The corresponding calculated theoretical value at $j_t=2.63$ is $\Omega_{G(2.63)\ th}=1.0043$ with a duality modulation of $+0.43\%$. The significance of the experimentally determined $\Omega_{G(2.63)\ ex}$ is most appropriately assessed relative to the no-grating control experiment value $\Omega_{NG\ ex}=0.9988\pm0.0018$. Since the theoretical prediction $\Omega_{G(2.63)\ th}>1$ and the theoretical prediction of the no-grating control experiment $\Omega_{NG\ th}=1$, the operant hypothesis is that the true mean of $\Omega_{G(2.63)\ ex}$ exceeds unity i.e. the true mean of $\Omega_{NG\ ex}$. Accordingly, the trials acquired for $\Omega_{G(2.63)\ ex}$ are statistically evaluated relative to the independent control experiment trials for $\Omega_{NG\ ex}$ in a one-tailed $t$ test to determine the $p$ confidence level. With the given experimental results, the calculated confidence level $p=0.015$ is highly supportive of the hypothesis that the true mean of $\Omega_{G(2.63)\ ex}>1$.

Similarly, with the 500 line/mm $G(3.16)$ grating installed, the value $\Omega_{G(3.16)\ ex}=0.9923\pm0.0019$ and the experimentally measured duality modulation is $-0.77\pm0.19\%$. $\Omega_{G(3.16)\ th}=0.9912$ giving a predicted duality modulation of $-0.88\%$. Assessing this result relative to $\Omega_{NG\ ex}$, the theoretical prediction $\Omega_{G(3.16)\ th}<1$ yields an operant hypothesis that the true mean of $\Omega_{G(3.16)\ ex}$ is less than unity. Applying the same statistical evaluation used for the previous grating, the calculated confidence level $p=0.011$ is highly supportive of the present hypothesis that the true mean of $\Omega_{G(3.16)\ ex}<1$.

Finally, the 400 line/mm grating $G(3.95)$ results in $\Omega_{G(3.95)\ ex}=0.9997\pm0.0023$ with a duality modulation of $-0.03\pm0.23\%$. Since $j_t=3.95$ is in the neighborhood of the $j_t=4$ diffraction null, there is no basis a priori for the true mean of $\Omega_{G(3.95)\ ex}$ to be statistically distinguishable from unity. This contention is supported by the theoretically predicted $\Omega_{G(3.95)\ th}=0.9985$ which closely approximates $\Omega_{NG\ th}=1$. Accordingly, a two-tailed $t$ test is appropriate for comparing the set of $\Omega_{G(3.95)\ ex}$ trials and the set of $\Omega_{NG\ ex}$ trials. The operant hypothesis then states that the true mean of $\Omega_{G(3.95)\ ex}$ is significantly different from that of $\Omega_{NG\ ex}$. However, with a calculated $p=0.76$, this hypothesis is strongly rejected and we conclude that the true means of $\Omega_{G(3.95)\ ex}$ and $\Omega_{NG\ ex}$ are not statistically distinguishable.

One additional statistical evaluation of interest can be performed with the above results. We have a theoretical basis for $\Omega_{G(2.63)\ th}>\Omega_{G(3.16)\ th}$. Consequently, we can propose a hypothesis that the true mean of $\Omega_{G(2.63)\ ex}$ exceeds that of $\Omega_{G(3.16)\ ex}$ and apply a one-tailed $t$ test to find the relevant $p$ confidence level. With a resultant $p=0.00016$, the present hypothesis is supported at an even higher confidence level than that for either $\Omega_{G(2.63)\ ex}$ or $\Omega_{G(3.16)\ ex}$ relative to $\Omega_{NG\ ex}$. Although this hypothesis does not provide a conclusion with respect to an absolute, i.e. that the true mean of $\Omega_{G(2.63)\ ex}>1$ or that the true mean of $\Omega_{G(3.16)\ ex}<1$, the confirmation of this hypothesis necessitates that at least one of the true means of $\Omega_{G(2.63)\ ex}$ and $\Omega_{G(3.16)\ ex}$ is not unity in violation of duality.

The above reported experimental results of $\Omega_{G\ ex}$ provide for the determination of the three points plotted on the theoretical graph of $\Omega_{G\ th}$ **Figure 4**. The particular gratings $G(2.63)$ and $G(3.16)$ are used because they are theoretically predicted to respectively produce enriched and depleted beams at $\lambda=633$ nm while $G(3.95)$ is predicted to produce an ordinary beam at that wavelength. Moreover, in the interests of facilitating replication of the present experiment, all three gratings are commercially readily available.

A detailed verification of the **Figure 4** theoretical curve by the acquisition of a large number of experimental data points is certainly of interest. However, that detailed verification is clearly not possible with three data points nor is that the present intent of the authors. The primary objective of the current experiment is to test for statistically significant violations of duality when particular resultant beams $\Phi_G$ are coupled with a beam $\Phi_R$. Specifically, $G(2.63)$ and $G(3.16)$ are predicted to prepare the resultant outgoing beams in a state not representable by duality. In the interest of eliminating potential systematic sources of experimental error, the respective resultant beams for all three gratings and for the NG control are maintained at the same power level for the complete set of trials. With this constraint of equivalent power, duality necessarily imposes a fundamental physical equivalence of the four resultant beams. Duality also forbids a net transfer for these or any other beams upon coupling with an independently generated beam. Nevertheless, when the four resultant beams are respectively coupled with $\Phi_R$, statistically significant net transfers consistently occur for $\Phi_{G(2.63)}$ and $\Phi_{G(3.16)}$ with positive and negative net transfers, respectively.

In addition to the trials reported here, many hundreds of preliminary trials have been conducted. Various beam powers for $\Phi_G$ and $\Phi_R$ have been employed for these trials while maintaining the general coupling criteria. Several different gratings of each of the three types were used. Throughout these preliminary trials the results were consistent with those reported here. An independent apparatus has also been assembled that similarly demonstrates $\Omega_{G\ ex}$ comparable to those of the present apparatus.

## 6. Discussion

From a theoretical perspective there remains an important consideration relating to the demonstration of duality violation as a validation of local realism. The probabilistic interpretation is distinguished from local realism as a consequence of non-locality manifested not only through duality but also through entanglement of correlated quantum entities [25]. The duality violation presented here is consistent with local realism. However, that leaves the dilemma of entanglement, seemingly conclusively confirmed through Bell's theorem [26] by reported experimental results [27-28], in support of the probabilistic interpretation despite compelling arguments to the contrary [29-30]. Any viable physical representation must necessarily demonstrate a self-consistent basis supporting or refuting both duality and entanglement.

In this regard, one of the present authors derived a locally real representation of quantum mechanical states from fundamental principles that gives agreement with performed experiments for correlated photons and for correlated particles [31]. That representation, demonstrating locality for correlated entities, is based on non-conservation of probability for individual states of those entities in accord with the present locally real representation demonstrating duality violation through non-conservation of probability in the diffraction process. Non-conservation of probability is of particular relevance in the present context. Clauser and Horne deduce that the locally real representations that can be excluded by Bell's theorem are constrained by an implicit supplementary assumption of "no-enhancement" (for which insertion of a polarizer does not increase detection) [32]. That exclusion does not apply to the locally real representation in [31] where non-conservation of probability inherently provides for enhancement. The resultant locally real representation is fully consistent with the underlying mathematical formalism of quantum mechanics. That formalism is "completed" [25] in the sense of maintaining local realism for quantum phenomena by admitting the degree of freedom to treat the relevant wave functions as relative entities of the field separable from resident particle-like entities. With this degree of freedom, specific examples of probability non-conservation emerge naturally.

Experimentally, the configuration reported here, consisting of a single beam incident on a conventional Ronchi grating, provides for a modest duality modulation. Nevertheless, even this modest duality modulation translates to a readily measurable increment of beam power in the macroscopic cw regime. Moreover, the associated underlying locally real basis for this configuration provides for a particularly straightforward and compelling understanding of a duality-violating phenomenon. There are, however, other configurations, e.g. [21], that exceed the present modest duality modulation. Ultimately, there is no inherent limitation on duality modulation in local realism.

The utility of a duality modulation with very high enrichment or depletion at macroscopic powers can be appreciated from the transient coupling used in the present configuration. Transient coupling shows that interaction of a duality-modulated beam with an ordinary beam results in a significant net transfer of energy from one of the beams to the other in the process. For example, a highly enriched beam can be used to directly amplify a weak ordinary "signal" beam. Transient coupling equilibrates the two beams causing an enrichment of the signal beam that enhances conventional detectability of the signal's wave modulations. Similarly, empty wave beams (or at least highly depleted beams) of macroscopic wave intensity would also be of utility in various applications such as probes of material samples where energy deposition into those samples by a probe beam must be minimized or eliminated. The interaction of an essentially empty probe beam with a sample would be made observable by equilibrating the post-interaction probe beam with an ordinary beam. That equilibration renders the wave of the post-interaction probe beam measurable by conventional detectors as a macroscopic power.

## 7. Conclusions

The experiment reported here demonstrates violations of quantum duality with an apparatus reduced to the simplest of elements. A laser beam is prepared by passage through one of three particular Ronchi gratings. When the prepared beam is coupled to an independent laser beam, the prepared beam transfers either +0.59% or −0.77% of its power to the independent beam for two of the gratings, respectively, whereas the power transfer associated with the prepared beam from the third grating is ≈0%. Quantum duality requires that the power transfer must be 0% for all three.

Notably, the experiment is conducted with beams of macroscopic power. The resultant duality-violating power transfers, representing extraordinarily large numbers of $\sim 10^{11}$ photons, are readily measurable by a conventional detector. The transfer phenomenon is robust and highly reproducible.

That transfer phenomenon is shown to be theoretically representable from basic principles and the phenomenon is consistent with recent reports [10] and [11] of other performed experiments that also demonstrate violations of duality.